\documentclass[sigconf]{acmart}

\usepackage{amsmath}
\usepackage{mathtools}
\usepackage{microtype}
\usepackage{graphicx}
\usepackage{booktabs}
\usepackage{multirow}
\usepackage{algorithm}
\usepackage{algorithmic}
\usepackage[table]{xcolor}
\usepackage{fontawesome5}
\usepackage{balance}

\definecolor{tablegray}{RGB}{245,245,245}

\AtBeginDocument{%
  }

\copyrightyear{2026}
\acmYear{2026}
\setcopyright{cc}
\setcctype{by}
\acmConference[KDD '26]{Proceedings of the 32nd ACM SIGKDD Conference on Knowledge Discovery and Data Mining V.2}{August 09--13, 2026}{Jeju Island, Republic of Korea}
\acmBooktitle{Proceedings of the 32nd ACM SIGKDD Conference on Knowledge Discovery and Data Mining V.2 (KDD '26), August 09--13, 2026, Jeju Island, Republic of Korea}
\acmDOI{10.1145/3770855.3818893}
\acmISBN{979-8-4007-2259-2/2026/08}
\begin{document}

\title{Ocean-E2E: Hybrid Physics-Based and Data-Driven Global Forecasting of Marine Heatwaves with End-to-End Neural Assimilation}

\author{Ruiqi Shu}
\email{srq24@mails.tsinghua.edu.cn}
\affiliation{%
  \institution{Department of Earth System Science, Tsinghua University}
  \city{Beijing}
  \country{China}
}

\author{Ruijian Gou}
\email{ruijian.gou@awi.de}
\affiliation{%
  \institution{Department of Ocean Big Data and Artificial Intelligence, Laoshan Laboratory}
  \city{Qingdao}
  \state{Shandong}
  \country{China}
}

\author{Yanfei Xiang}
\email{xiangyf22@mails.tsinghua.edu.cn}
\affiliation{%
  \institution{Department of Earth System Science, Tsinghua University}
  \city{Beijing}
  \country{China}
}

\author{Xiaomeng Huang}
\authornote{Corresponding author.}
\email{hxm@tsinghua.edu.cn}
\affiliation{%
  \institution{Department of Earth System Science, Tsinghua University}
  \city{Beijing}
  \country{China}
}

\renewcommand{\shorttitle}{%
  \begin{tabular}[t]{@{}l@{}}
    Ocean-E2E: Hybrid Physics-Based and Data-Driven Global Forecasting\\
    of Marine Heatwaves with End-to-End Neural Assimilation
  \end{tabular}}
\renewcommand{\shortauthors}{Ruiqi Shu, Ruijian Gou, Yanfei Xiang, and Xiaomeng Huang}

\begin{abstract}
This work focuses on the end-to-end forecast of global extreme marine heatwaves (MHWs), which are unusually warm sea surface temperature events with profound impacts on marine ecosystems. Accurate prediction of extreme MHWs has significant scientific and financial worth. However, existing methods still have certain limitations in forecasting general patterns and extreme events. In this study, to address these issues, based on the physical nature of MHWs, we created a novel hybrid data-driven and numerical MHWs forecast framework Ocean-E2E, which is capable of 40-day accurate MHW forecasting with end-to-end data assimilation. Our framework significantly improves the forecast ability of MHWs by explicitly modeling the effect of oceanic mesoscale advection and air-sea interaction based on a dynamic kernel. Furthermore, Ocean-E2E is capable of end-to-end MHWs forecast and regional high-resolution prediction, allowing our framework to operate completely independently of numerical models while outperforming the current state-of-the-art ocean numerical/AI forecasting-assimilation models. Experimental results show that the proposed framework performs excellently on global-to-regional scales and short-to-long-term forecasts, especially in those most extreme MHWs. Overall, our model provides a framework for forecasting and understanding MHWs and other climate extremes. The source code is available at \url{https://github.com/ChiyodaMomo01/Ocean-E2E}.

\end{abstract}



\maketitle

\section{Introduction}

Marine heatwaves (MHWs) are abnormally warm seawater events that significantly damage marine ecosystems \cite{oliver2021marine}. In other words, MHWs are extreme \textbf{Sea Surface Temperature Anomaly (SSTA)} events. MHWs, particularly extreme ones, can cause coral bleaching \cite{hughes2017global, hughes2018spatial} and widespread mortality of marine organisms \cite{garrabou2009mass, thomson2015extreme}. As a result, precise forecasting of extreme MHWs has significant scientific and economic implications. For example, synoptic scale MHWs forecasting can help seafood production and management planning, such as feed cycles, at 1-7 day timescales,  while subseasonal to seasonal forecasting can further support proactive decision-making for the blue economy \cite{hobday2016seasonal, malick2020environmentally, mills2017forecasting, payne2022skilful}. In this study, we will mainly focus on the subseasonal-to-seasonal forecast (i.e., 1-40 days).\par

Traditional MHW forecasting relies mainly on two paradigms: physics-based numerical models and emerging data-driven approaches. Numerical models solve oceanic primitive equations to produce seasonal forecasts that capture large-scale patterns of MHW onset and intensification \cite{jacox2022global,brodie2023ecological}, as well as sub-seasonal predictions that resolve finer-scale variability over shorter lead times \cite{benthuysen2021subseasonal,yu2024assessing}. In parallel, recent advances in deep learning have shown promise for efficient global ocean forecasting \cite{cui2025forecasting,wang2024xihe,xiong2023ai,huang2025fuxi}, with extensions to MHW prediction \cite{lin2023rossby,sun2023artificial}. Despite their operational value, both paradigms suffer from inherent limitations:

1) \textbf{Numerical models}: These approaches demand substantial computational resources \cite{xiong2023ai,wang2024xihe}, especially in the data assimilation processes. Moreover, they often parameterize key physical processes—such as air-sea coupling and mesoscale eddy dynamics—through empirical formulations, leading to systematic biases and constrained skill in predicting the general patterns of MHWs \cite{jacox2022global,giamalaki2022assessing}.

2) \textbf{Data-driven models}: While these methods statistically approximate complex oceanic dynamics, they frequently generate non-physical fields, exhibiting limited accuracy in forecasting extreme events like MHWs due to issues such as error accumulation and smoothing of SST gradients during autoregressive rollouts \cite{wang2024xihe}. Additionally, they typically depend on initialization from numerical models, precluding end-to-end operation from raw observational inputs \cite{xiong2023ai}.

Given these complementary strengths and weaknesses, a natural question arises: \textit{why not integrate the advantages of both by combining a relatively simple physical model with AI components, where the physical model handles well-understood dynamics and AI simulates the more complex or uncertain aspects?} In this study, we pursue this hybrid strategy to effectively mitigate the aforementioned drawbacks of numerical and data-driven models, enabling our framework to harness the best of both worlds:

1) \textbf{Enhanced computational efficiency and end-to-end capability}: By substituting neural networks for the forecasting and data assimilation processes, we have substantially reduced the cost of predicting MHWs, enabling our framework to operate independently of numerical models.

2) \textbf{Improved physical realism and reduced biases}: The incorporation of physical constraints ensures generation of consistent fields, addressing the non-physical artifacts and extreme-event limitations in pure data-driven methods.

To address these integrated shortcomings, we introduce Ocean-E2E, a global hybrid physics-based data-driven framework that improves forecasts of extreme MHWs, drawing inspiration from numerical models. The contributions of this paper can be summarized as follows: (1) \textbf{Global/Regional MHWs Forecasting Framework}. We propose a hybrid physics-based data-driven method that supports both global scale and regional high-resolution forecasting, achieving high-accuracy results for extreme MHWs forecasts. (2) \textbf{Global MHWs Neural Assimilation Framework}. We designed an end-to-end MHWs neural system that, when combined with the forecasting model, can directly obtain the complete initial field and forecasting results of MHWs from observation data. (3) \textbf{Physics-consistent Forecast}. By combining physical laws into our framework, our model has achieved state-of-the-art results in long-term and regional high-resolution forecasting with higher physical consistency.


\begin{figure*}[t]
\centering
\includegraphics[width=1\linewidth]{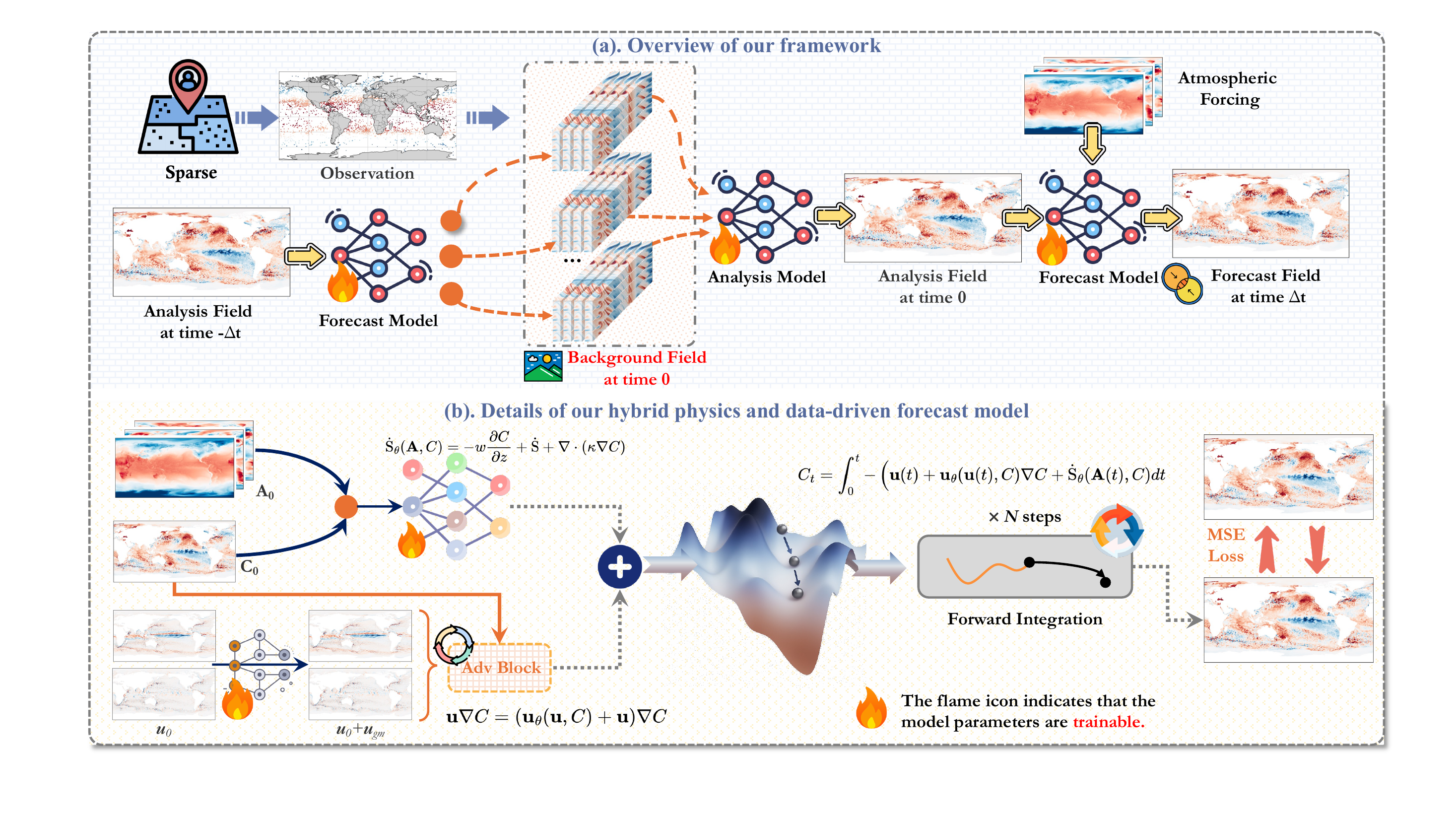}
\caption{Our proposed Ocean-E2E framework. a) Overview of our framework. b) Details of our hybrid physics and data-driven forecast model.}
\Description{A two-part schematic showing the end-to-end observation-to-forecast pipeline and the hybrid physics-data-driven forecast model for marine heatwave prediction.}
\label{fig:visual}
\end{figure*}

\section{Related Work}
The method most similar to us is ClimODE \cite{verma2024climode}. Similarly, it also uses advection partial differential equations to design the forecasting framework. However, our work has many differences and improvements compared to ClimODE: (1) \textbf{Differences in theoretical framework.} From the perspective of the theoretical framework, in their work, the `speed' of the flow field is essentially an abstract mathematical definition, whose core function is to maintain the conservation of passive tracers. In our work, the flow field speed strictly corresponds to the realistic ocean flow field and models the additional transport effect of the flow field on passive tracers at sub-grid scales through mesoscale `bolus' velocity. (2) \textbf{Differences in model design.} ClimODE uses an emission model to simulate the unclosed and uncertain portions of the real system, while our model further relates this part to the ocean-atmosphere interaction and is designed based on physical theory to form the structure of the source term. (3) \textbf{Differences in application.} ClimODE requires solving the initial velocity field as the initial condition for the model, which involves solving a physics-informed neural network (PINNs) for each training sample. This makes the ClimODE framework difficult to scale up. What's more, our experiments (Section \ref{ocean_simulation}) show that our method achieves greater stability and accuracy in MHWs forecasting compared to ClimODE.
\section{Method}
\subsection{Framework Overview}
In this study, the problem of the end-to-end MHWs forecast is \textbf{essentially forecasting SSTA}, which can be separated into two parts (shown in Figure \ref{fig:visual}). First, estimating the current state of SSTA $C_{0} \in \mathbb{R}^{H \times W}$ (a global two-dimensional field, where $H$ and $W$ correspond to latitude and longitude) based on the observation $C_{0}^{obs}$ and the background field $C_{0}^{bg}$. Second, predicting the future state of mixed layer SSTA $C_{t}$ based on the current state $C_{0}$ under atmospheric forcing $A_{0:t} \in \mathbb{R}^{C \times H \times W}$. Here $A_{0:t}$ denotes the atmosphere state from time $0$ to $t$. Formally speaking, we are going to model the conditional distribution $p(C_{t}|C_{0}^{obs}, C_{0}^{bg}, A_{0:t},)$ in two stages:
\begin{equation}
    p(C_t|C_0^{obs}, C_0^{bg}, A_{0:t}) = \underbrace{p^f(C_t|C_0, A_{0:t})}_{\text{Forecast model}} \cdot \underbrace{p^a(C_0|C_0^{obs}, C_0^{bg})}_{\text{Assimilation model}}
\end{equation}
where $p^{f}$ and $p^{a}$ denote forecast and assimilation process, respectively. In our end-to-end framework, we model these two processes through two specifically designed neural networks $\phi^{f}_{\theta}$ and $\phi^{a}_{\theta}$, which we will discuss in detail in the following sections.
\subsection{\texorpdfstring{The Design of $\phi^{f}_{\theta}$: Hybrid Physics-AI Model}{The Design of the Forecast Model: Hybrid Physics-AI Model}}
\label{subsec:forecast_design}
As established in the problem formulation, forecasting MHWs fundamentally reduces to predicting extreme sea surface temperature anomalies (SSTA). To effectively embed physical principles into our modeling framework, we must rigorously account for the governing laws of SSTA evolution. We approach this challenge through a generalized geophysical fluid dynamics perspective, wherein seawater temperature constitutes a \textit{passive tracer} advected by oceanic flow fields. Formally speaking, the governing equation of SSTA $C$, together with other surface variables can be written as: \par

\begin{align}
\frac{\partial C}{\partial t} + ({\bf u}_g + {\bf u}_{ag}) \nabla C +w \frac{\partial C}{\partial z} & = \dot{\rm S} \label{eq:1} \\
f\mathbf{k} \times \mathbf{u}_g & = - \nabla (g \eta) \label{eq:2} \\
\frac{\partial \eta}{\partial t} + \nabla \cdot (H\hat{\mathbf{u}}) & = 0 \label{eq:3} \\
\frac{\partial \hat{\mathbf{u}}}{\partial t} + \widehat{\mathbf{G}_u} + \nabla (g \eta) & = 0 \label{eq:4} \\
w | _{z=0} & = \eta_{t} \label{eq:5}
\end{align}

Where $t$ is time, $({\bf u}, w)=(\mathbf{u}_{g}+\mathbf{u}_{ag}, w)$ are horizontal and vertical flow velocity ($\mathbf{u}_g$ represents oceanic geostrophic velocity while $\mathbf{u}_{ag}$ is ageostrophic velocity), $\nabla$ is the horizontal derivation, $\dot{\rm S}$ is the sink/source, $\mathbf{k}$ is a unit vector in the vertical direction, $f$ is the Coriolis parameter, $\eta$ is sea surface height, $H$ is ocean depth, $\hat{\cdot}$ is vertical average operator (i.e., $\hat{\mathbf{u}} = \int_{-H}^{0} \mathbf{u} dz$), $g$ is the gravitional accerelation. By integrating equation (1), the state of $C$ at time $t$ can be expressed in the form of the initial condition $C_{0}$:
\begin{equation}
\begin{split}
C_{t} &= \int_{0}^{t} -{\bf u}_g \nabla C + (-{\bf u}_{ag} \nabla C-w \frac{\partial C}{\partial z} + \dot{\rm S}) \, dt + C_0 \\
&\simeq \int_{0}^{t} \underbrace{-{\bf u}_g \nabla C}_{\text{advective transport}} + \underbrace{(-{\bf u}_{ag} \nabla C + \dot{\rm S})}_{\text{mixing and external sink/source}} \, dt + C_0
\end{split}
\label{evo_of_c}
\end{equation}

The second equal sign is due to the fact that, according to equation (5), $w|_{z=0}$ is small ($10^{-6} \sim 10^{-5}$ m/s) compared to other terms. Equation (\ref{evo_of_c}) implies two key governing mechanisms of MHWs: 1) \textbf{\textit{advective transport}} by geostrophic currents that redistribute thermal energy, 2) \textbf{\textit{mixing and external sink/source}} encompassing convective mixing and external ageostrophic forcing through surface boundary interactions. In the following paragraphs, we will combine physical theory and deep learning methods to simulate these two mechanisms in four steps.\par

\textbf{\textit{Step 1. Modelling advective transport}}. The key factor of advective transport is the ocean surface current ${\bf u}_g$, where the flow field at the smaller scale (mesoscale) plays an important role in the distribution of MHWs and other oceanic passive tracers. However, in most models and data, due to limited resolution, the model cannot distinguish these small-scale processes, so the advection processes caused by the flow field below the model's finest grid (i.e., subgrid processes):
\begin{equation}
\overline{{\bf u}_g \nabla C} = \overline{{\bf u}_g} \overline{\nabla C} + \overline{{\bf u}_g^{\prime}{\nabla C}^{\prime}}
\end{equation}
are not fully simulated. Here, the overline $\overline{\cdot}$ represents the downsampling operator. In other words, the variables with an overline denote the low-resolution data. In the traditional numerical model, Gent and McWilliams \cite{gent1990isopycnal} found that this subgrid process can be fitted by `bolus velocity' (also called GM90 parameterization): 
\begin{equation}
\overline{{\bf u}_g^{\prime}{\nabla C}^{\prime}} \sim {\bf u}_{gm}\overline{\nabla C}
\end{equation}

According to the GM90 parameterization, these bolus velocity can be expressed as \cite{gent1990isopycnal}:
\begin{equation}
{\bf u}_{gm} = (\kappa \nabla\rho/\rho_z)_z
\end{equation}
where $\rho = \rho(C)$ is the density. It should be noticed that this expression requires subsurface ocean data, which is often extremely difficult to obtain. What's more, the choice of coefficient $\kappa$ requires an empirical formula, which is always inaccurate. Alternatively, we seek a neural network ${\bf u}_{\theta}({\bf u}_g, C) = {\bf u}_{gm}$ to approximate the bolus velocity and the subgrid advection effect. In summary, the advection term can be expressed as:
\begin{equation}
{\bf u}_g \nabla C = ({\bf u}_{\theta}({\bf u}_g, C) + {\bf u}_g) \nabla C
\end{equation}

\textbf{\textit{Step 2. Modelling mixing and external sink/source}}. This term is strongly related to the air-sea interaction along the ocean's surface. To be specific, for SSTA, the source term consists of four parts:
\begin{equation}
\dot{\rm S} \propto Q_{{\rm net}}^{\prime} = Q_{{\rm SW}}^{\prime}+Q_{{\rm LW}}^{\prime}+Q_{{\rm sh}}^{\prime}+Q_{{\rm lh}}^{\prime}
\end{equation}
where $Q_{{\rm SW}}, Q_{{\rm LW}}, Q_{{\rm sh}}, Q_{{\rm lh}}$ are the shortwave radiation, longwave thermal radiation, sensible heat flux, and latent heat flux, respectively. Of the four air-sea heat fluxes, the sensible and latent flux $Q_{{\rm sh}}^{\prime}+Q_{{\rm lh}}^{\prime}$ determines the majority of the SSTA's variations \cite{holbrook2019global}. According to the bulk formula \cite{cui2025forecasting}, these two terms are significantly related to surface wind speed $U_{10}, V_{10}$, near-surface temperature of the atmosphere $T_{2m}$, and surface specific humidity $q_a$. What's more, the ageostrophic velocity $\mathbf{u}_{ag}$ is alse driven by surface wind forcing $U_{10}$, $V_{10}$. Based on the observation above, we assume that mixing and external sink/source can be well approximated by the surface variables of the atmosphere ${\bf A} = (U_{10},V_{10},T_{2m},q_{a})$ and $C$ itself. Specifically,
\begin{equation}
\dot{\rm S}_{\theta}({\bf A},C) = -{\bf u}_{ag} \nabla C + \dot{\rm S}
\end{equation}

Since this study primarily focuses on proposing a physics-AI \textbf{framework}, the design and selection of the specific structures for the two neural networks ${\bf u}_{\theta}, \dot{\rm S}_{\theta}$ are highly flexible. In this study, we selected a powerful and robust spatiotemporal prediction backbone \cite{gao2022simvp} to accomplish this task. We summarize the architecture of our backbone as below \cite{gao2022simvp}:

\textbf{Spatial Encoding.} The model begins with the spatial encoder, which extracts latent features frame-by-frame using strided convolutional blocks (ConvSC). This stage compresses the spatial resolution while expanding the channel dimension to capture high-level physical abstractions, mathematically expressed as $\mathbf{Z} = \mathcal{E}_{enc}(\mathbf{X}_{in})$, where $\mathbf{Z} \in \mathbb{R}^{T \times C' \times H' \times W'}$ represents the compressed latent states sequence.

\textbf{Spatiotemporal Translation.} The core evolution dynamics are modeled by the translator module (implemented as `Mid\_Xnet'), which employs a U-Net-like structure populated with Inception blocks. To capture multi-scale spatiotemporal dependencies without recurrent computation, the translator aggregates features using a diverse set of kernels $\mathcal{K} = \{3, 5, 7, 11\}$. For a given hidden feature $\mathbf{h}$, the Inception operation performs a multi-branch aggregation: $\mathbf{h}_{out} = \sum_{k \in \mathcal{K}} \sigma \left( \text{GroupConv}_{k \times k}(\mathbf{h}_{in}) \right)$, where $\sigma$ denotes the LeakyReLU activation and GroupConv ensures parameter efficiency.

\textbf{Spatial Decoding.} Finally, the evolved latent features are reconstructed back to the original physical domain via the decoder. This module utilizes transposed convolutions to upsample the spatial resolution, yielding the predicted next-step physical states: $\hat{\mathbf{Y}} = \mathcal{D}_{dec}(\mathbf{Z}_{evolved})$, where the output $\hat{\mathbf{Y}}$ matches the dimensions of the input $\mathbf{X}_{in}$, completing the end-to-end prediction loop.
We believe that more advanced backbones can further enhance the model's prediction performance.

\textbf{\textit{Step 3. Treating boundary conditions}}. Above paragraphs point out that in order to effectively model the key physical mechanism of MHWs, we need the future state of $\mathbf{A}$ and $\mathbf{u}_g$ (i.e., boundary conditions). For oceanic geostrophic velocity $\mathbf{u}_g$, $\frac{\partial}{\partial t} (3) - \nabla \cdot (H (4))$ implies that 
\begin{equation}
\frac{\partial^2 \eta}{\partial t^2} + \nabla \cdot (gH \nabla \eta) = \nabla \cdot \widehat{\mathbf{G}_u} \simeq \nabla \cdot \widehat{\mathbf{G}_{u_{g}}} \triangleq \mathcal{G}(\eta) 
\end{equation}
which is a self-consistent hyperbolic equation. This indicates that, at least at a relatively short time interval, the evolution of $\eta$, together with geostrophic velocity, can be modelled based on an auto-regressive manner:
\begin{equation}
\frac{\partial {\bf u_g}}{\partial t} = \mathcal{M}_{\theta}({\bf u_g})
\end{equation}
The neural network $\mathcal{M}_{\theta}$ adopts a U-Net-like encoder-decoder architecture specifically optimized for ocean surface geostrophic velocity modeling. The core computational unit is the \textbf{Group Attention Block (\texttt{GABlock})}, which is designed to capture multi-scale spatio-temporal dependencies efficiently.

The \texttt{GABlock} consists of two sequential sub-modules: a Spatial Attention (\texttt{SA}) module and a Multilayer Perceptron (\texttt{Mlp}).
Given an input feature $\mathbf{X}$, the \texttt{SA} module first expands the receptive field using a cascade of depth-wise ($\text{DWConv}$) and depth-wise dilated convolutions ($\text{DWDConv}$). The intermediate features are then split into content $\mathbf{F}$ and a gating signal $\mathbf{G}$ to perform an element-wise attention operation:
\begin{equation}
\begin{aligned}
    [\mathbf{F}, \mathbf{G}] &= \text{Split}(\text{DWDConv}(\text{DWConv}(\mathbf{X}))), \\
    \mathbf{Y}_{sa} &= \text{Proj}\left( \mathbf{F} \odot \sigma(\mathbf{G}) \right),
\end{aligned}
\end{equation}
where $\sigma(\cdot)$ denotes the Sigmoid activation. The output of the \texttt{SA} module is stabilized via a residual connection with a learnable scale $\lambda_1$, yielding $\mathbf{X}_{sa} = \mathbf{X} + \lambda_1 \cdot \mathbf{Y}_{sa}$.

Subsequently, the \texttt{Mlp} module refines channel-wise representations. It employs a $\text{DWConv}$ sandwiched between two $1\times1$ convolutions to aggregate local context. The final output of the \texttt{GABlock} is obtained through a second scaled residual connection:
\begin{equation}
    \mathbf{X}_{out} = \mathbf{X}_{sa} + \lambda_2 \cdot \text{Mlp}(\text{BN}(\mathbf{X}_{sa})).
\end{equation}
This hierarchical design allows $\mathcal{M}_{\theta}$ to effectively model both fine-grained turbulent eddies and large-scale circulation patterns.

What's more, the atmosphere variables between time interval $[0, t]$ is also acquired through a large AI-driven weather forecast model \cite{gao2025oneforecast}:
\begin{equation}
\frac{\partial {\bf A}}{\partial t} = \mathcal{N}_{\theta}({\bf A})
\end{equation}

\textbf{\textit{Step 4. Combining physics and AI together}}. As shown in the section above, the furture state of $C$ at time $t$ can be modelled as:
\begin{equation}
\begin{split}
C_{t} = &\int_{0}^{t} -\Bigg[\int_{0}^{s} \mathcal{M}_{\theta}({\bf u}_g)ds + {\bf u}_{\theta}(\int_{0}^{s} \mathcal{M}_{\theta}({\bf u}_g)ds, C )\Bigg ] \nabla C \\
&+ \dot{\rm S}_{\theta}(\int_{0}^{s} \mathcal{N}_{\theta}({\bf A})ds,C)  ds
\end{split}
\end{equation}

In the training stage, we first pretrain these two neural networks $\mathcal{M}_{\theta}, \mathcal{N_{\theta}}$. Then the parameters of these networks are frozen and we utilize the forecast model to optimize the parameters of ${\bf u}_{\theta}, \dot{\rm S}_{\theta}$ based on the MSE loss, which can be written as:
\begin{equation}
\mathcal{L} = ||\hat{C_{t}} - \int_{0}^{t} -({\bf u}_g + {\bf u}_{\theta}({\bf u}_g, C)) \nabla C  + \dot{\rm S}_{\theta}(\mathbf{A},C)  dt ||^2
\label{loss}
\end{equation}
Where $\hat{C_{t}}$ is the groundtruth data of $C_t$. In practice, by separating the time interval $[0,t]$ into $N$ subintervals $[i \Delta t, (i+1)\Delta t], 1 \leq i \leq N-1$, equation (\ref{loss}) is approximated via forward Euler method: 

\begin{align}
C_{(i + 1)\Delta t} &= C_{i\Delta t} + \Big[-\big(\mathbf{u}_{i\Delta t} + \mathbf{u}_{\theta}(\mathbf{u}_{i\Delta t}, C_{i\Delta t})\big) \nabla C_{i\Delta t} \notag \\
&\quad + \dot{\rm S}_{\theta}(\mathbf{A}_{i\Delta t},\mathbf{A}_{(i + 1)\Delta t},C_{i\Delta t})\Big]\Delta t, \\
\mathbf{u}_{(i+1) \Delta t} &= \mathbf{u}_{i \Delta t}+\mathcal{M}_{\theta}(\mathbf{u}_{i \Delta t})\Delta t, \\
\mathbf{A}_{(i+1) \Delta t} &= \mathbf{A}_{i \Delta t}+\mathcal{N}_{\theta}(\mathbf{A}_{i \Delta t})\Delta t.
\end{align}
Then the parameters of ${\bf u}_{\theta}, \dot{\rm S}_{\theta}$ are optimized through $\mathcal{L}=||\hat{C_{t}} - C_{N\Delta t}(\theta)||^2$. 
It is worth noting that the above numerical kernel is highly \textbf{lightweight}, imposing no significant computational burden on our framework. Details of the numerical implementation are summarized in Appendix~\ref{app:reproducibility}.

\subsection{\texorpdfstring{The Design of $\phi^{a}_{\theta}$: Neural Data Assimilation}{The Design of the Assimilation Model: Neural Data Assimilation}}

As established in the previous section, the hybrid framework relies on an accurate initial condition $C_0$. However, in operational scenarios, the full analysis field is unobservable. This necessitates a data assimilation (DA) process to fuse sparse observations $C_0^{obs}$ with a background estimate $C_0^{bg}$.

We first generate the background field $C_0^{bg}$ by perturbing the state at the previous time step $C_{-\Delta t}$ with structured noise within the assimilation window $[-\Delta t, 0]$:
\begin{equation}
C_0^{bg} = \phi^{f}_{\theta}(C_{-\Delta t} + \varepsilon, \mathbf{A}_{0:\Delta t}),
\end{equation}
where $\varepsilon$ denotes Perlin noise. The core innovation of our framework lies in the neural assimilation network, $\phi^{a}_{\theta}$, which establishes a direct, non-linear mapping to produce the final analysis field:
\begin{equation}
C_{0} = \phi^{a}_{\theta}(C_{0}^{obs}, C_{0}^{bg}).
\end{equation}
This initialization supports a continuous autoregressive correction loop, where the assimilation network rectifies the forecast trajectory at each step $t$:
\begin{equation}
C_{t+\Delta t} = \phi^{a}_{\theta}(C_{t+\Delta t}^{obs}, \phi^{f}_{\theta}(C_{t} + \varepsilon_t, \mathbf{A}_{t:t+\Delta t})).
\end{equation}

The network $\phi^{a}_{\theta}$ is an edge-reparameterization and attention-guided architecture designed for real-time projection of sparse observations. As detailed below, it integrates a Kirsch-guided Reparameterization Module (KRM), dual-domain Attention Mechanisms, and a cascaded backbone.

\subsubsection{Kirsch-guided Reparameterization Module (KRM)}
To efficiently capture sharp SST boundaries, we employ the KRM \cite{liu2024real}, which utilizes a multi-branch training structure incorporating standard convolutions and a specialized edge-detection branch based on eight-directional Kirsch filters.

Crucially, for inference, these branches are merged into a single $3\times3$ convolution via structural reparameterization. The equivalent parameters are derived as:
\begin{equation}
\begin{aligned}
    W_{\mathrm{rep}} &= W_{std} + W_{es} + \sum_{i=1}^{8} \left( \mathrm{perm}(W_i) * W_{\mathrm{K}}^i \right), \\
    B_{\mathrm{rep}} &= B_{std} + B_{es} + \sum_{i=1}^{8} \left( \mathrm{perm}(B_i) * B_{\mathrm{K}_i} \right),
\end{aligned}
\end{equation}
where $W_{\mathrm{K}}^i$ and $B_{\mathrm{K}_i}$ represent the scaled Kirsch filters and biases, and $\mathrm{perm}(\cdot)$ denotes dimensional permutation. This design ensures that the network retains explicit edge-awareness without incurring any additional computational overhead during deployment.

\subsubsection{Joint Spatial-Channel Attention}
To handle the sparsity of $C^{obs}$, we employ a joint attention mechanism.
The \textbf{Channel Attention Module (CAM)} recalibrates feature responses to emphasize informative channels:
\begin{equation}
    \operatorname{CAM}(x) = \sigma\Big(\mathrm{MLP}\big(\mathrm{AvgPool}(x)\big) + \mathrm{MLP}\big(\mathrm{MaxPool}(x)\big)\Big),
\end{equation}
where $\sigma$ is the Sigmoid function. Simultaneously, the \textbf{Spatial Attention Module (SAM)} focuses on crucial spatial regions (e.g., observing stations vs. missing areas):
\begin{equation}
    \operatorname{SAM}(x) = \sigma\Big(\mathrm{Conv}_{7\times7}\big([\mathrm{AvgPool}(x) ; \mathrm{MaxPool}(x)]\big)\Big),
\end{equation}
where $[\cdot;\cdot]$ denotes channel-wise concatenation.

\subsubsection{Cascaded Assimilation Backbone}
$\phi^{a}_{\theta}$ utilizes a residual cascade to progressively refine the fused input into a high-fidelity analysis field $C_0$. Implementation details are summarized in Appendix~\ref{app:reproducibility}.

\section{Experiments}
We designed comprehensive experiments to evaluate our model. Our evaluation is guided by the following key questions:
\textbf{RQ1:} \textbf{Overall Performance:} Can Ocean-E2E outperform state-of-the-art models with higher consistency and better performance in normal and extreme MHW events?
\textbf{RQ2:} \textbf{End-to-end Forecasting:} Can Ocean-E2E run independently of numerical model while preserving a high accuracy in real-world MHWs forecast?
\textbf{RQ3:} \textbf{High-resolution Multiscale Prediction:} Can our framework achieve a satisfying performance in regional high-resolution simulations, where complicated oceanic multi-scale dynamics pose more challenges to the model's forecast skill and physical consistency?
\subsection{Benchmarks and Baselines}
We conduct the experiments on GLORYS12V1 reanalysis data \cite{lellouche2018copernicus}, which provides daily mean data of sea surface temperature (SST) covering latitudes between -80° and 90° spanning between 1993 and 2021. The seasonal cycle has been removed from the original data in order to get the MHWs (SSTA) field \cite{shu2025advanced}. The subset we use includes years from 1993 to 2021, which is 1993-2018 for training, 2019 for validating, and 2020-2021 for testing. The surface velocity of the ocean is obtained from satellite observation \cite{pujol2016duacs}. The atmospheric variables are obtained from ECMWF Reanalysis v5 (ERA5) \cite{hersbach2020era5}. Furthermore, the observational data employed for assimilation encompasses satellite-derived ocean observations, denoted as $C^{obs}_{sat}$, and sparse in-situ measurements, $C^{obs}_{sparse}$. Further details regarding data sources and preprocessing procedures are provided in Appendix~\ref{app:reproducibility}.

\begin{table*}[t]
\caption{In the global ocean simulation task, we compare the performance of our Ocean-E2E with various baselines. The average results for global SSTA (Sea Surface Temperature Anomaly) are measured using RMSE and CSI. Lower RMSE ($\downarrow$) and higher CSI ($\uparrow$) indicate better performance. The best results are in \textbf{bold}, and the second-best are \underline{underlined}. Note that some baseline models are unstable and produce unrealistic forecasts, with RMSE exceeding 10 K replaced by ---.}
\label{tab:simulation_res}
\vskip 0.1in
\centering

\renewcommand{\multirowsetup}{\centering}


\begin{tabular*}{0.9\textwidth}{@{\extracolsep{\fill}}l@{\hspace{5pt}}l|cc|cc|cc|cc|cc}
\toprule

\multicolumn{2}{l|}{\multirow{3}{*}{Model Category}} & \multicolumn{10}{c}{Metric} \\
\cmidrule(lr){3-12}
\multicolumn{2}{l|}{} & \multicolumn{2}{c}{20-day} & \multicolumn{2}{c}{30-day} & \multicolumn{2}{c}{40-day} & \multicolumn{2}{c}{50-day} & \multicolumn{2}{c}{60-day} \\
\cmidrule(lr){3-12}
\multicolumn{2}{l|}{} & RMSE & CSI & RMSE & CSI & RMSE & CSI & RMSE & CSI & RMSE & CSI \\
\midrule

\rowcolor{tablegray}
\multicolumn{12}{l}{\textbf{Ocean/Weather Forecasting Models}} \\ 
\faCloud & FourCastNet \cite{pathak2022FourCastNet} & 0.6836 & 0.2709 & 0.9852 & 0.2161 & --- & 0.1755 & --- & 0.1537 & --- & 0.1423 \\
\faCloud & CirT \cite{liu2025cirt} & 1.3496 & 0.0905 & 1.6838 & 0.0852 & 1.9249 & 0.0810 & 1.9932 & 0.0770 & 2.0321 & 0.0746 \\
\faCloud & WenHai \cite{cui2025forecasting} & \textbf{0.5435} & \underline{0.4202} & \underline{0.6277} & \underline{0.3414} & \underline{0.7006} & \underline{0.2805} & \underline{0.7633} & \underline{0.2447} & \underline{0.8139} & \underline{0.2209} \\
\faCloud & ClimODE \cite{verma2024climode} & 0.7221 & 0.3263 & 0.7963 & 0.2594 & 0.8555 & 0.2091 & 0.8997 & 0.1896 & --- & 0.1754 \\
\midrule

\rowcolor{tablegray}
\multicolumn{12}{l}{\textbf{Operator Learning Models}} \\
\faPuzzlePiece & CNO \cite{raonic2023convolutional} & 0.7062 & 0.3669 & 0.8091 & 0.2938 & 0.9025 & 0.2367 & 0.9821 & 0.2011 & 1.0556 & 0.1771 \\
\faPuzzlePiece & LSM \cite{wu2023solving} & 1.1346 & 0.3025 & 1.5512 & 0.2291 & --- & 0.1760 & --- & 0.1436 & --- & 0.1312 \\
\midrule

\rowcolor{tablegray}
\multicolumn{12}{l}{\textbf{Computer Vision Backbones}} \\
\faCameraRetro & U-Net \cite{ronneberger2015u} & 0.6923 & 0.3909 & 0.7952 & 0.3191 & 0.8893 & 0.2629 & 0.9706 & 0.2292 & 1.0463 & 0.2054 \\
\faCameraRetro & ResNet \cite{he2016deep} & --- & 0.2537 & --- & 0.1983 & --- & 0.1538 & --- & 0.1248 & --- & 0.1029 \\
\faCameraRetro & DiT \cite{peebles2023scalable} & 0.9390 & 0.3411 & 1.2649 & 0.2919 & 1.7051 & 0.2528 & 2.2474 & 0.2311 & 2.9139 & 0.2137 \\
\midrule

\rowcolor{tablegray}
\multicolumn{12}{l}{\textbf{Spatiotemporal Models}} \\
\faFilm & ConvLSTM \cite{shi2015convolutional} & 0.7135 & 0.3585 & 0.8040 & 0.2856 & 0.8920 & 0.2292 & 0.9726 & 0.1957 & 1.0545 & 0.1735 \\
\faFilm & SimVP \cite{gao2022simvp} & 0.6729 & 0.3749 & 0.7609 & 0.3193 & 0.8345 & 0.2736 & 0.8864 & 0.2413 & 0.9296 & 0.2162 \\
\faFilm & PastNet \cite{wu2024pastnet} & 1.3876 & 0.1867 & 1.4073 & 0.1809 & 1.4230 & 0.1760 & 1.4287 & 0.1733 & 1.4353 & 0.1705 \\
\midrule

\faTrophy & \textbf{Ocean-E2E} & \underline{0.5659} & \textbf{0.4285} & \textbf{0.6163} & \textbf{0.3712} & \textbf{0.6596} & \textbf{0.3230} & \textbf{0.6911} & \textbf{0.2874} & \textbf{0.7194} & \textbf{0.2580} \\
& Promotion & --- & 2.0\% & 1.8\% & 8.7\% & 5.8\% & 15.1\% & 9.5\% & 17.4\% & 11.6\% & 16.8\% \\

\bottomrule
\end{tabular*}
\end{table*}

\begin{figure}[t]
\centering
\includegraphics[width=\linewidth]{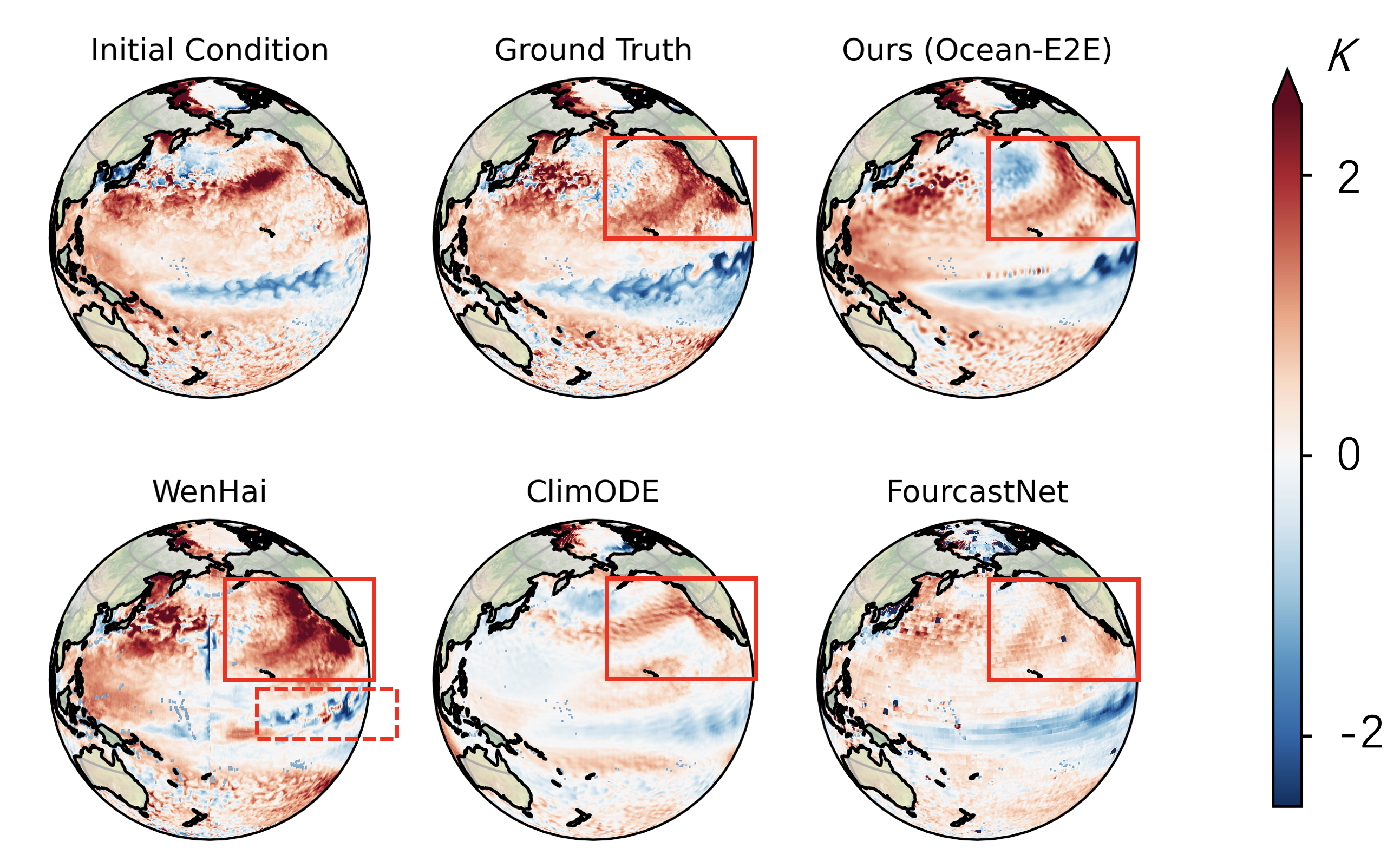}
\caption{Snapshots of our framework and other baseline simulations at 60-day lead time. The simulations are initialized on August 7th, 2020. Red solid boxes represent MHW events while red dashed box indicates physical inconsistency in forecasting results.}
\Description{Global maps comparing Ocean-E2E and baseline simulations at a 60-day lead time, with boxes marking marine heatwave events and physically inconsistent predictions.}
\label{fig:snapshot}
\end{figure}

\begin{table*}[t]
\caption{In operational global ocean forecast task, we compare the performance of our Ocean-E2E with S2S. The average results for global SSTA of RMSE are recorded for 2020 and 2021. A small RMSE ($\downarrow$) indicates better performance. The best results are in \textbf{bold}.}
\label{tab:forecast_res}
\centering

\begin{sc}
\renewcommand{\multirowsetup}{\centering}


\begin{tabular*}{0.8\textwidth}{@{\extracolsep{\fill}}l|cccc|cccc}
\toprule
\multirow{3}{*}{Model} & \multicolumn{8}{c}{Metric (RMSE)} \\
\cmidrule(lr){2-9}
& \multicolumn{4}{c}{2020} & \multicolumn{4}{c}{2021} \\
\cmidrule(lr){2-5} \cmidrule(lr){6-9}
& 10-day & 20-day & 30-day & 40-day & 10-day & 20-day & 30-day & 40-day \\
\midrule
S2S & 0.8140 & 0.8870 & 0.9514 & 0.9965 & 0.8261 & 0.8897 & 0.9450 & 0.9895 \\
Ocean-E2E (Ours) & \textbf{0.5750} & \textbf{0.6047} & \textbf{0.7514} & \textbf{0.8747} & \textbf{0.6414} & \textbf{0.7447} & \textbf{0.8147} & \textbf{0.8729} \\
\midrule
Ocean-E2E (Promotion) & 29.4\% & 31.8\% & 21.02\% & 12.2\% & 22.4\% & 16.3\% & 13.7\% & 11.7\% \\
\bottomrule
\end{tabular*}
\end{sc}
\end{table*}

\begin{figure*}[t] 
  \centering
  \includegraphics[width=0.8\textwidth]{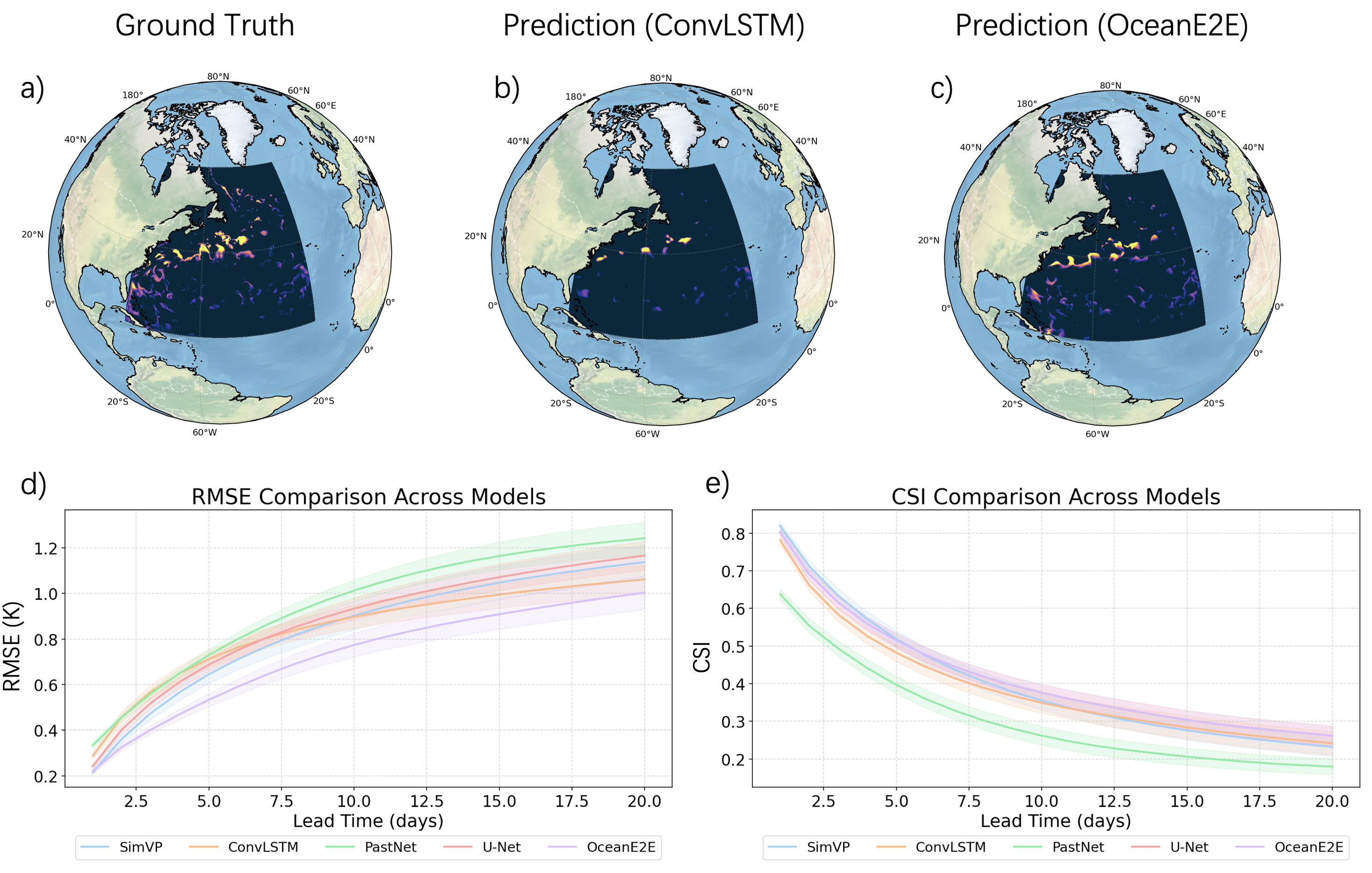}
  \caption{Results of our regional high resolution MHWs simulation in western Atlantic Ocean. a)-c) Snapshots of our simulation, where \textbf{yellow} parts indicate MHWs. d)-e) RMSE and CSI of our Ocean-E2E and other baselines.}
  \Description{Regional western Atlantic maps and metric curves comparing Ocean-E2E with baselines for high-resolution marine heatwave simulation.}
  \label{fig:highres}
\end{figure*}

Before introducing our experiments, there are two key concepts that should be clarified: \textbf{ocean simulation} and \textbf{end-to-end ocean forecast}. As pointed out in previous studies \cite{shu2025advanced, cui2025forecasting}, the ocean simulation utilizes the current state of SSTA $C_0$ to forecast future state $C_t$ under the realistic atmosphere forcing ${\bf A}_{0:t}$. In contrast, the end-to-end forecast task includes assimilating the initial analysis field of SSTA $C_0$ and forecasting only based on the current atmosphere state ${\bf A}_{0}$.

We compare the simulation performance of Ocean-E2E with 4 main categories of data-driven models: \textbf{ocean forecasting models} (FourCastNet \cite{kurth2023fourcastnet}, CirT \cite{liu2025cirt}, WenHai \cite{cui2025forecasting} and ClimODE \cite{verma2024climode}), \textbf{operator learning models} (FNO \cite{li2020fourier}, CNO \cite{raonic2023convolutional} and
LSM \cite{wu2023solving}), \textbf{computer vision backbones} (U-Net \cite{ronneberger2015u}, ResNet \cite{he2016deep} and DiT \cite{peebles2023scalable}) and \textbf{spatiotemporal models} (ConvLSTM \cite{shi2015convolutional}, SimVP \cite{gao2022simvp} and PastNet \cite{wu2024pastnet}).

We also compare the end-to-end operational forecast ability against \textbf{traditional numerical operational forecast system}: ECMWF sub-seasonal to seasonal prediction (S2S). S2S is currently the superior subseasonal-to-seasonal weather forecast model that provides the forecast of atmosphere and ocean surface physics at a lead time of 40 days. The original resolution is 1.5°, and we upsample the data to 1/2° for comparison.

\subsection{Evaluation Metric}
We utilized two metrics, \textbf{RMSE} (Root Mean Square Error) and \textbf{CSI} (Critical Success Index), to evaluate the forecast performance. The RMSE represents the \textbf{overall performance} of our framework's forecast (lower is better), while CSI mainly focuses on the most \textbf{extreme MHW events} (higher is better). What's more, we utilize assimilation bias (BIAS) to assess the systematic bias induced by our neural assimilation system. Metric definitions are summarized in Appendix~\ref{app:reproducibility}.

\subsection{Evaluation of ocean simulation performance (RQ1)}
\label{ocean_simulation}
 As shown in Table \ref{tab:simulation_res}, it can be found that in long-term ocean simulations (40-60 days), our framework performs better compared to the baseline. On day 60, the RMSE decreased by an average of 11.6$\%$ and the CSI increased by an average of 9.2$\%$. In shorter simulations (20 days), our model was slightly lower than WenHai but significantly higher than other baselines. This indicates that our Ocean-E2E can capture both general patterns and extreme events of MHWs. To further illustrate our simulation ability, we select a snapshot of our 60 days simulation, as shown in Figure \ref{fig:snapshot}. It can be found that a strong MHW is growing in the West Pacific Ocean (red solid box in Figure \ref{fig:snapshot}). Compared to other baselines, our Ocean-E2E shows a robust forecast result and better consistency with ground truth.

\subsection{Evaluation of the end-to-end neural assimilation and forecast performance (RQ2)}
\label{subsec:e2e_task}
We conducted two sets of assimilation experiments using analysis fields from January 1, 2020 and January 1, 2021 as initial conditions, respectively. Both experiments employed a 6-day assimilation window to assimilate SSTA fields over a full year. As shown in Appendix~\ref{app:additional_results} (Figure~\ref{fig:assimilation}), the RMSE and BIAS evolution reveal that assimilation errors rapidly increase and stabilize during the assimilation process, demonstrating the stability of our assimilation framework.

Using these assimilated fields as initial conditions and GLORYS12V1 reanalysis as ground truth, we evaluate the forecasting performance of Ocean-E2E against the S2S system (shown in Table \ref{tab:forecast_res}). \textbf{Since S2S provides 46 days of forecast data, in this study, we only compare the 10-40 days' forecast results of Ocean-E2E with it}. To ensure experimental fairness, results in Table \ref{tab:forecast_res} were calculated after the assimilation outputs stabilized (12 days post-initialization). It can be found that Ocean-E2E achieves an average 10$\%$ reduction in RMSE compared to superior end-to-end numerical prediction systems across 40-day subseasonal-to-seasonal forecasts. Additional comparative experiments using S2S assimilation fields as both reference truth and initial conditions show the same trend (Table~\ref{tab:forecast_s2s_init}). In all experimental configurations, Ocean-E2E demonstrates superior performance relative to S2S, confirming its robust competitiveness.


\subsection{Regional high-resolution simulation (RQ3)}
As discussed in Section 2.1, our framework builds upon the GM90 theoretical framework, which is also applicable to high-resolution data. To validate its performance, we implement a regional high-resolution simulation ($1/12^\circ$) of the western Atlantic Ocean. Figure~\ref{fig:highres} a)-c) illustrate a comparison between Ocean-E2E and baselines. Compared to the ground truth, our model has more physical details with more extreme MHW events (yellow parts in Figure~\ref{fig:highres} a)). Evaluation based on RMSE and CSI (Figure~\ref{fig:highres} d)-e)) also confirmed that Ocean-E2E achieves a state-of-the-art performance in regional simulations. Furthermore, to verify physical consistency, we calculated the power spectrum of the SSTA gradient $|\nabla C|^2$ based on the Ocean-E2E simulation results and compared it with the baselines (Appendix~\ref{subapp:pd}). The results demonstrate that our model exhibits higher consistency with the real physical fields, especially at small scales (high frequency/wavenumber).

\begin{table}[t]
  \centering
  \small
  \setlength{\tabcolsep}{3pt} 
  
  \caption{Comprehensive ablation studies. \textbf{Panel A} analyzes the impact of physical design, forcings, and architecture on simulation accuracy (CSI $\uparrow$). \textbf{Panel B} evaluates the assimilation efficacy and error propagation (RMSE $\downarrow$). New experiments in Panel B isolate the contributions of observational data types and neural components. The results of our original Ocean-E2E model are marked in bold.}
  \label{tab:ablation_combined}
  
  \begin{tabular}{lcccc}
    \toprule
    \multicolumn{5}{l}{\textbf{Panel A: Component \& Architecture Analysis (Metric: CSI)}} \\
    \addlinespace[3pt]
    
    \multirow{2}{*}{\textit{Model Variants}} & \multicolumn{2}{c}{Global Simulation} & \multicolumn{2}{c}{Regional Simulation} \\
    \cmidrule(lr){2-3} \cmidrule(lr){4-5}
     & 20-day & 60-day & 10-day & 30-day \\
    \midrule
    
    \multicolumn{5}{l}{\textit{Physical Model Design}} \\
    \hspace{1em} $\dot{S}_{\theta}$ (S only) & 0.3884 & 0.1658 & 0.3913 & 0.1877 \\
    \hspace{1em} $(\mathbf{u}_{\theta}+\mathbf{u}_g) \nabla C$ (ADV only) & 0.3207 & 0.1504 & 0.3207 & 0.0963 \\
    \hspace{1em} $\mathbf{u}_g \nabla C$ (Pure Numerical) & --- & --- & --- & --- \\
    \addlinespace[4pt] 
    
    \multicolumn{5}{l}{\textit{Time-Varying Forcings}} \\
    \hspace{1em} Static Atmosphere ($\mathbf{A}(0)$) & 0.2652 & 0.1223 & 0.3021 & 0.1470 \\
    \hspace{1em} Static Background (${\bf u}_g(0)$) & 0.3988 & 0.2115 & 0.4080 & 0.2245 \\
    \addlinespace[4pt]
    
    \multicolumn{5}{l}{\textit{Neural Architecture}} \\
    \hspace{1em} w/o \texttt{GABlock} & 0.4015 & 0.2245 & 0.4322 & 0.2588 \\
    \hspace{1em} w/o \texttt{MidXnet} & 0.3540 & 0.1810 & 0.3890 & 0.2015 \\
    \addlinespace[2pt]
    
    \rowcolor{gray!10} 
    \hspace{1em} \textbf{Ocean-E2E (Full Model)} & \textbf{0.4285} & \textbf{0.2580} & \textbf{0.4615} & \textbf{0.2926} \\

    \midrule
    \addlinespace[6pt] 
    
    \multicolumn{5}{l}{\textbf{Panel B: Assimilation \& Error Propagation (Metric: RMSE)}} \\
    \addlinespace[3pt]
    
    \multirow{2}{*}{\textit{Assimilation Variants}} & \multicolumn{2}{c}{Year 2020} & \multicolumn{2}{c}{Year 2021} \\
    \cmidrule(lr){2-3} \cmidrule(lr){4-5}
     & 20-day & 40-day & 20-day & 40-day \\
    \midrule
    
    \multicolumn{5}{l}{\textit{Input Data Source Analysis}} \\
    \hspace{1em} w/o Sparse Obs ($C^{obs}_{sparse}$) & 0.6215 & 0.8901 & 0.7580 & 0.8850 \\
    \hspace{1em} w/o Satellite Obs ($C^{obs}_{sat}$) & 0.6850 & 0.9520 & 0.8105 & 0.9410 \\
    \hspace{1em} w/o $\mathcal{N}_{\theta}$ (Use Real Atm) & 0.5944 & 0.8325 & 0.7297 & 0.8132 \\
    \hspace{1em} w/o $\phi_{\theta}^{a}$ (Use Real IC) & 0.5832 & 0.8691 & 0.6965 & 0.8531 \\
    \addlinespace[4pt]
    
    \multicolumn{5}{l}{\textit{Assimilation Network Components}} \\
    \hspace{1em} w/o KRM Module & 0.6350 & 0.8955 & 0.7620 & 0.8910 \\
    \hspace{1em} w/o Attention (CAM/SAM) & 0.6280 & 0.8890 & 0.7590 & 0.8875 \\
    \addlinespace[2pt]
    
    \rowcolor{gray!10}
    \hspace{1em} \textbf{Ocean-E2E (Full Model)} & \textbf{0.6047} & \textbf{0.8747} & \textbf{0.7447} & \textbf{0.8729} \\
    
    \bottomrule
  \end{tabular}
\end{table}

\subsection{Ablation studies}

To explicitly validate the effectiveness and robustness of the proposed framework, we conducted comprehensive ablation studies presented in Table~\ref{tab:ablation_combined}. We disentangle the contributions of core components across three critical dimensions: \textbf{physical model design}, \textbf{neural architecture with dynamic forcings}, and the \textbf{data assimilation strategy}.

\paragraph{Physical Model Design (Panel A)}
We first evaluate the two fundamental terms in the SSTA evolution equation: the advection term $({\bf u}_{\theta} + {\bf u}_g) \nabla C$ (ADV) and the source term $\dot{S}_{\theta}$ (S). As shown in the first section of Panel A, removing either component leads to a marked reduction in Critical Success Index (CSI). Notably, the source term $\dot{S}_{\theta}$ contributes most significantly to forecasting performance, underscoring the dominance of air-sea thermodynamic interaction in MHW evolution. Furthermore, a pure numerical baseline (removing the neural velocity correction ${\bf u}_{\theta}$) results in model collapse due to numerical instability caused by grid-scale noises. This highlights the necessity of the learned `bolus' velocity, which effectively acts as a subgrid damping term to stabilize the physical system.

\paragraph{Time-Varying Forcings and Architecture (Panel A)}
We further investigated the impact of dynamic environments and specific neural modules. 
First, regarding environmental forcings, fixing the atmosphere ($\mathbf{A}$) or background flow (${\bf u}_g$) to their initial states (static) causes a significant performance drop (e.g., Global 60-day CSI decreases from 0.2580 to 0.1223). This confirms that capturing the time-varying nature of heat flux and large-scale ocean circulation is essential for long-term simulation. 
Second, regarding neural architecture, removing the Group Attention Block (\texttt{GABlock}, core design of $\mathcal{M}_{\theta}$ as mentioned in Section \ref{subsec:forecast_design}) or the \texttt{MidXnet} (core design of ${\bf u}_{\theta}$ and $\dot{\rm S}_{\theta}$, which are also mentioned in Section \ref{subsec:forecast_design}) leads to a substantial decline in accuracy. This validates their respective roles in encoding background physics and capturing complex spatiotemporal dynamics in the latent space.

\paragraph{Assimilation Data Source and Components (Panel B)}
To understand the efficacy of our end-to-end assimilation strategy, we analyzed the impact of different data sources and neural components within the assimilation network $\phi_{\theta}^{a}$ based on the end-to-end forecast task (see Section \ref{subsec:e2e_task}). In terms of \textbf{input data}, removing satellite observations ($C^{obs}_{sat}$) results in a sharp increase in RMSE compared to removing sparse observations ($C^{obs}_{sparse}$), reflecting the critical role of high-coverage satellite data in constraining the global field.
In terms of \textbf{network design}, ablating the Kirsch-guided Reparameterization Module (w/o KRM) or the Attention mechanism (w/o Attn) both lead to higher reconstruction errors. This suggests that simply fusing data is insufficient; the network must explicitly capture sharp thermal boundaries (via KRM) and focus on data-rich regions (via Attention) to generate high-quality analysis fields.

\paragraph{Systemic Error Propagation (Panel B)}
Finally, we present the ablation results for the core components of the marine heatwave forecasting task to investigate how errors propagate through the atmospheric forcing forecast module $\mathcal{N}_{\theta}$ and the assimilation module $\phi_{\theta}^{a}$. As shown in Table \ref{tab:ablation_combined}, comparing the full Ocean-E2E model with variants using real atmospheric forcing (w/o $\mathcal{N}_{\theta}$) and real initial conditions (w/o $\phi_{\theta}^{a}$), we observe that the error in the atmospheric forecast module accumulates gradually over time (e.g., the gap between w/o $\mathcal{N}_{\theta}$ and Full model widens at 40-day lead time). This underscores the critical role of accurate atmospheric forcing fields. In contrast, the assimilation module does not exhibit a significant cumulative error effect, as it primarily impacts the state fields in the immediate vicinity of the assimilation timestamp, thereby exerting a localized effect.

\section{Conclusions}

In this study, we introduce Ocean-E2E, a hybrid framework merging data-driven and numerical methods for end-to-end marine heatwave (MHW) forecasting. By resolving mesoscale advection and air-sea interactions with a dynamic kernel, it effectively bridges the gap between physical interpretability and computational efficiency, delivering stable 40-day global-to-regional MHW predictions in both simulation and forecast tasks. Crucially, neural assimilation enables independent global forecasting with enhanced accuracy and stability. Comprehensive results confirm its robustness across scales, advancing MHW prediction and providing critical support for understanding the multi-scale evolution of extreme ocean climate events.

\section{Limitations and Ethical Considerations}
We list several limitations and ethical considerations of our work here:

\textbf{Ethical Considerations:} This study focuses on developing a hybrid physics-informed AI framework for forecasting marine environmental events. To the best of our knowledge, this work does not raise significant ethical or privacy concerns.

\textbf{Limitations:} Despite the superior performance of Ocean-E2E, we acknowledge two potential limitations. First, as a data-driven approach, inherent biases or missing values in the ground-truth data could potentially influence fine-grained local predictions. Second, while the model demonstrates robust generalization on historical data, its extrapolation capability under unprecedented future climate scenarios remains an open question for the broader AI4Science community and requires further long-term monitoring.

\section{GenAI Disclosure}
During the preparation of this work, the authors used Generative AI in order to improve the language and readability. After using this tool, the authors reviewed and edited the content and take full responsibility for the content of the manuscript.
\begin{acks}
This work was supported by the National Natural Science Foundation of China (42125503, 42430602).
\end{acks}
\bibliographystyle{ACM-Reference-Format}
\balance
\bibliography{sample-base}

\appendix
\section{Reproducibility and Implementation Details}
\label{app:reproducibility}

\textbf{Algorithm.}
Algorithm~\ref{alg:oceane2e} summarizes the end-to-end assimilation and forecast loop. At each assimilation time, $N=10$ Perlin-noise perturbations are added to the previous analysis state, and the forecast model propagates these perturbed states over the assimilation window to generate an ensemble of background fields. The neural assimilation network then fuses the background ensemble with sparse and satellite observations to obtain the analysis field. Starting from this assimilated state, the forecast model advances SSTA using predicted atmospheric forcing, geostrophic velocity, learned bolus-velocity correction, and learned source terms.

\begin{algorithm}[h]
\caption{Ocean-E2E Framework for Global MHW Forecast}
\label{alg:oceane2e}
\begin{algorithmic}[1]
\renewcommand{\algorithmicrequire}{\textbf{Require:}}
\REQUIRE SSTA observations $C_t^{obs}$, previous analysis field $C_{t-1}^{a}$, atmospheric forcing $\mathbf{A}_t$, and ocean velocity $\mathbf{O}_t$.
\ENSURE Analysis field $C_t^a$ and next-step forecast $C_{t+1}$.
\REPEAT
\STATE Generate background ensembles with $N=10$ Perlin-noise perturbations: $C_t^{bg(i)}=\phi^f_\theta(C_{t-1}^{a}+\varepsilon^{(i)},\mathbf{A}_{t-1:t})$.
\STATE Fuse observations and background fields with the neural assimilation network: $C_t^a=\phi^a_\theta(C_t^{bg},C_t^{obs})$.
\STATE Predict boundary conditions: $\mathbf{A}_{t+1}=\mathcal{N}_\theta(\mathbf{A}_t)$ and $\mathbf{O}_{t+1}=\mathcal{M}_\theta(\mathbf{O}_t)$.
\STATE Integrate the hybrid physics-data-driven forecast kernel to obtain $C_{t+1}$, and set $t\leftarrow t+1$ for the next assimilation/forecast step.
\UNTIL the desired forecast horizon is reached.
\RETURN $C_t^a,C_{t+1}$.
\end{algorithmic}
\end{algorithm}

\textbf{Data summary.}
\begingroup\sloppy
The GLORYS12V1 SST/SSTA data used in the main experiments span 1993--2021: 1993--2018 are used for training, 2019 for validation, and 2020--2021 for testing. The SST target is selected from GLORYS12V1, downloaded from \url{https://data.marine.copernicus.eu/product/GLOBAL_MULTIYEAR_PHY_001_030}. Following the main experiments, we remove the seasonal cycle from the raw SST field to obtain the SSTA field used for MHW forecasting.

For the atmospheric forcing module $\mathcal{N}_{\theta}$, we utilize OneForecast~\cite{gao2025oneforecast}. The data used to train OneForecast are obtained from WeatherBench2 benchmark data derived from ERA5 reanalysis~\cite{rasp2024weatherbench,hersbach2020era5}. The atmospheric fields include 13-level geopotential, specific humidity, temperature, and horizontal winds, together with surface 10 m u-component of wind (U10M), 10 m v-component of wind (V10M), 2 m temperature (T2M), and mean sea level pressure (MSLP). These variables are provided on a $1.5^\circ$ global grid at 6-hour resolution. For the ocean-current module $\mathcal{M}_{\theta}$, we use DUACS satellite-altimetry surface geostrophic velocities, including eastward and northward components $(U_g,V_g)$, at $0.25^\circ$ daily resolution, downloaded from \url{https://data.marine.copernicus.eu/product/SEALEVEL_GLO_PHY_L4_MY_008_047}.

For the hybrid forecast kernel, the global experiments use 10 m u-component of wind (U10M), 10 m v-component of wind (V10M), 2 m temperature (T2M), $(U_g,V_g)$, raw SST fields, and the derived SSTA fields over 1993--2021. The global products are aligned to the common $1/2^\circ$ grid used in the simulation and forecast comparisons. For the regional high-resolution experiment, we use the western Atlantic subset ($20^\circ$--$60^\circ$N, $30^\circ$--$80^\circ$W), where the same atmospheric, velocity, SST, and SSTA fields are evaluated at $1/12^\circ$ resolution. For observational data, we utilize Global Ocean ODYSSEA L4 Sea Surface Temperature (satellite observations), \url{https://data.marine.copernicus.eu/product/SST_GLO_PHY_L4_MY_010_044}, and the Hadley Centre Integrated Ocean Database (sparse observations), \url{https://climatedataguide.ucar.edu/climate-data/hadiod-met-office-hadley-centre-integrated-ocean-database}.
\par\endgroup

\textbf{Preprocessing.}
To preserve the scale of the advective terms, ocean variables (SSTA, $U_g$, and $V_g$) are kept in physical units and are not normalized. Instead, normalization is only applied to the atmospheric variables used by the hybrid forecast kernel, including U10M, V10M, and T2M. The mean $\mu$ and standard deviation $\sigma$ are estimated from the 1993--2018 training split, and each raw atmospheric variable $\mathbf{X}$ is transformed as $\mathbf{X}'=(\mathbf{X}-\mu)/\sigma$.

The data sources differ in both temporal and spatial resolution. For temporal alignment, daily averages and daily snapshots are combined at the daily scale without additional temporal interpolation. For spatial alignment, we use bilinear interpolation, implemented with \texttt{torch.nn.functional.interpolate}, to map all variables to a unified $1/2^\circ$ grid for global simulations and a $1/12^\circ$ grid for regional simulations. Sparse in-situ observations are first filtered to sea-surface records. We then construct a $1/2^\circ$ global grid and assign each observation to grid cells within a $1.5^\circ \times 1.5^\circ$ window centered on the observation point.

\textbf{Numerical integration and efficiency.}
\begingroup
\setlength{\abovedisplayskip}{3pt}
\setlength{\belowdisplayskip}{3pt}
\setlength{\abovedisplayshortskip}{2pt}
\setlength{\belowdisplayshortskip}{2pt}
As shown in Section~\ref{subsec:forecast_design}, the evolution equation of the hybrid physics-data-driven framework can be written as
\begin{equation}
\label{eq:appendix_evo_core}
C_{t} = \int_{0}^{t} \left\{ -({\bf u}_g(t) + {\bf u}_{\theta}({\bf u}_g(t), C)) \nabla C  + \dot{\rm S}_{\theta}({\bf A}(t),C) \right\} dt + C_0.
\end{equation}
For spatial discretization, we use second-order centered finite differences to approximate $\nabla C$:
\begin{equation}
\nabla_h C(i,j)=\big(({\bf D_x}*C)_{i,j},({\bf D_y}*C)_{i,j}\big),
\end{equation}
where $*$ is the convolution operator, and ${\bf D_x}$ and ${\bf D_y}$ are horizontal-gradient stencils:
\begin{equation}
{\bf D_x} = \begin{pmatrix}
0 & 0 & 0 \\
-1/d_{i,j} & 0 & 1/d_{i,j} \\
0 & 0 & 0
\end{pmatrix},
{\bf D_y} = \begin{pmatrix}
0 & 1/s_{i,j} & 0 \\
0 & 0 & 0 \\
0 & -1/s_{i,j} & 0
\end{pmatrix}.
\end{equation}
Here $d_{i,j}$ and $s_{i,j}$ are the longitudinal and latitudinal distances between the two neighboring cells used by the centered stencil at $(i,j)$. Unlike atmospheric variables, ocean MHW fields have coastline boundaries, which should be treated properly to avoid numerical instability. We mitigate this boundary effect through a boundary mask $\mathbf{M}$:
\begin{equation}
\mathbf{M}(i,j)=
\begin{cases}
0, & \text{if the grid cell is situated on land,}\\
   & \text{or if any of its neighboring cells are on land},\\
1, & \text{otherwise}.
\end{cases}
\end{equation}
Furthermore, a scaling factor $\varepsilon_{gm}=0.1$ is employed on ${\bf u}_{\theta}$ to avoid instability during training. Equation~\ref{eq:appendix_evo_core} then becomes
\begin{equation}
\begin{split}
C_{t} &= \int_{0}^{t} \left\{ \left[ -\left( {\bf u}_g(t) + \varepsilon_{gm}{\bf u}_{\theta}({\bf u}_g(t), C) \right) \nabla C \right]\mathbf{M} \right. \\
&\quad \left. + \dot{\rm S}_{\theta}({\bf A}(t),C) \right\} dt + C_0.
\end{split}
\end{equation}
For compact notation in the discrete updates, define the one-step forecast tendency as
\begin{align}
\mathcal{F}(C,\mathbf{u},\mathbf{A},\mathbf{A}') &=
-\big(\mathbf{u}+\varepsilon_{gm}\mathbf{u}_{\theta}(\mathbf{u},C)\big)
\cdot\nabla_h C\odot \mathbf{M} \notag\\
&\quad+\dot{\rm S}_{\theta}(\mathbf{A},\mathbf{A}',C).
\label{eq:appendix_tendency}
\end{align}
We employ the forward Euler method to approximate the time integration of the system. The global $1/2^\circ$ simulation uses a 24-hour step and integrates each four-day forecast block with four explicit Euler updates:
\begin{align}
C^{n+1} &= C^n+\Delta t\,
\mathcal{F}(C^n,\mathbf{u}_g^n,\mathbf{A}^n,\mathbf{A}^{n+1}),
\label{eq:appendix_global_euler}\\
\mathbf{u}_g^{n+1}&=\mathbf{u}_g^n+\Delta t\,\mathcal{M}_{\theta}(\mathbf{u}_g^n),\\
\mathbf{A}^{n+1}&=\mathbf{A}^n+\Delta t\,\mathcal{N}_{\theta}(\mathbf{A}^n).
\end{align}
For the regional $1/12^\circ$ western Atlantic simulation, mesoscale eddies and submesoscale fronts evolve more rapidly, so we use $\Delta t=1800$ seconds and perform 192 sub-steps over the same four-day interval. To balance computational efficiency with fidelity, the neural corrections $\mathbf{u}_{\theta}$ and $\dot{\rm S}_{\theta}$ are applied every 48 small steps, corresponding to one day. For each daily block $k$, the geostrophic transport is first advanced for $j=0,\ldots,47$:
\begin{equation}
C_{k,j+1}=C_{k,j}-\Delta t\big(\mathbf{u}_{g,k,j}\cdot\nabla_h C_{k,j}\big)\odot \mathbf{M}.
\label{eq:appendix_regional_substep}
\end{equation}
At the end of the daily block, the neural corrections are applied once:
\begin{align}
\mathbf{u}_{g,k+1}&=\mathbf{u}_{g,k}+\Delta T\,\mathcal{M}_{\theta}(\mathbf{u}_{g,k}),\qquad
\mathbf{A}_{k+1}=\mathbf{A}_{k}+\Delta T\,\mathcal{N}_{\theta}(\mathbf{A}_{k}),\\
C_{k+1,0}&=C_{k,48}+\Delta T\,\mathcal{R}_{k},\\
\mathcal{R}_{k}&=
-\varepsilon_{gm}\mathbf{u}_{\theta}(\mathbf{u}_{g,k+1},C_{k,48})
\cdot\nabla_h C_{k,48}\odot \mathbf{M} \notag\\
&\quad+\dot{\rm S}_{\theta}(\mathbf{A}_{k},\mathbf{A}_{k+1},C_{k,48}),
\label{eq:appendix_daily_correction}
\end{align}
where $\Delta T=24$ hours. On a single NVIDIA RTX 3090 GPU with batch size 1, a 30-day forecast takes 195.29 ms with the dynamical kernel and 171.58 ms without it, indicating that the physics kernel adds only modest overhead.

\textbf{Evaluation protocol.}
Global forecasts are initialized from analysis fields and evaluated against GLORYS12V1 SSTA on the common $1/2^\circ$ grid. For S2S comparisons, all products are interpolated to the same grid, restricted to available operational lead times, and scored after the assimilation outputs stabilize, i.e., 12 days post-initialization. Regional experiments use the same metrics on the $1/12^\circ$ western Atlantic domain, where small-scale SST fronts and MHW boundaries pose greater challenges to forecast skill and physical consistency.

\textbf{Metrics.}
We identify MHWs following the Hobday definition: an event occurs when SST exceeds the local 90th percentile for at least five consecutive days. RMSE evaluates global SSTA accuracy, CSI evaluates extreme-event detection, and BIAS evaluates systematic assimilation error:
\begin{align}
{\rm RMSE}(t) &= \sqrt{\frac{1}{H W}\sum_{i,j}\omega_{i,j}(C_{i,j}(t)-\tilde{C}_{i,j}(t))^2},\\
{\rm CSI}(t) &= \frac{{\rm TP}}{{\rm TP}+{\rm FP}+{\rm FN}},\\
{\rm BIAS}(t) &= \frac{1}{H W}\sum_{i,j}\omega_{i,j}(C_{i,j}(t)-\tilde{C}_{i,j}(t)).
\end{align}
Here $C$ and $\tilde{C}$ denote the ground-truth and predicted SSTA fields, $\omega_{i,j}$ is the area-weighting coefficient, and $(i,j)$ indexes a grid cell in an $H\times W$ field. TP (True Positive) denotes the number of cases that a MHW event is accurately predicted. FP (False Positive), FN (False Negative), and TN (True Negative) follow a similar definition.
\endgroup

\textbf{Training.}
All baseline models and Ocean-E2E are trained under the same experimental protocol: 100 epochs, initial learning rate $10^{-3}$, and a step-wise learning-rate scheduler. Final checkpoints are selected by validation performance.

\section{Additional Evaluation Results}
\label{app:additional_results}

\begin{figure}[t]
\centering
\includegraphics[width=0.95\linewidth]{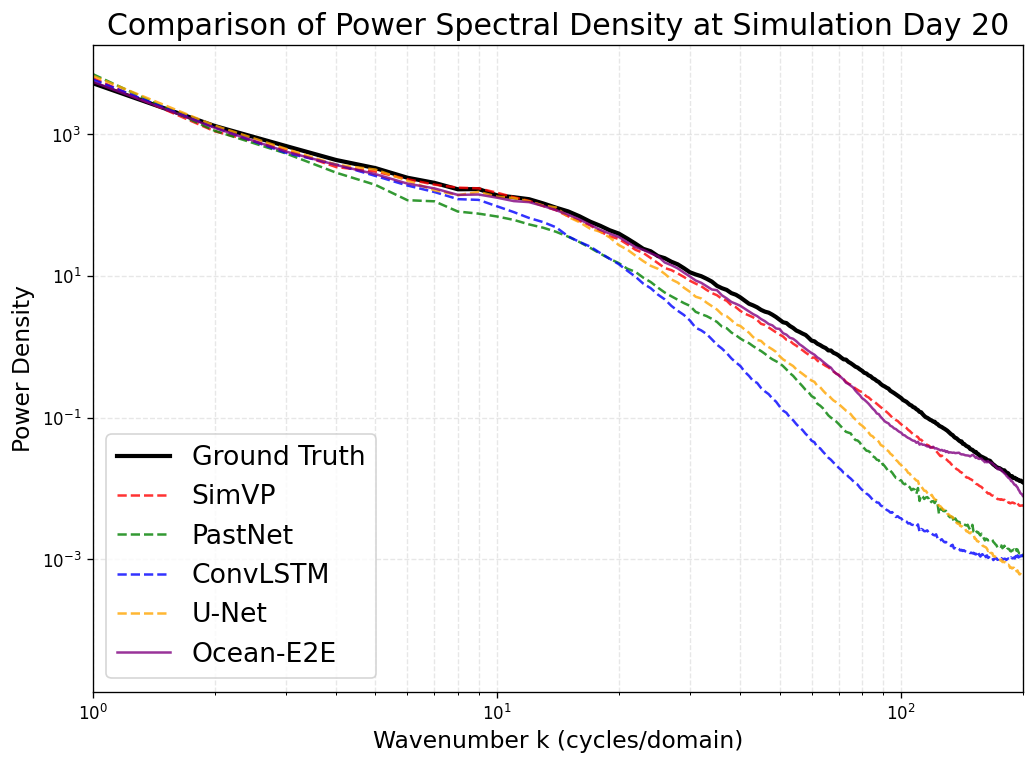}
\caption{Power spectrum of $|\nabla C|^2$ on day 20 from the regional simulation.}
\Description{A power-spectrum comparison showing that Ocean-E2E follows the ground-truth curve more closely than baseline models at high wavenumbers.}
\label{fig:pd}
\end{figure}

\textbf{Spectral physical consistency.}
\label{subapp:pd}
Figure~\ref{fig:pd} reports the power spectrum of the squared SSTA-gradient magnitude $|\nabla C|^2$ on day 20 of the regional simulation. Ocean-E2E is closer to the ground truth at larger wavenumbers, indicating that the model preserves more small-scale thermal-gradient structure than the baselines.

\begin{figure}[t]
\centering
\includegraphics[width=\linewidth]{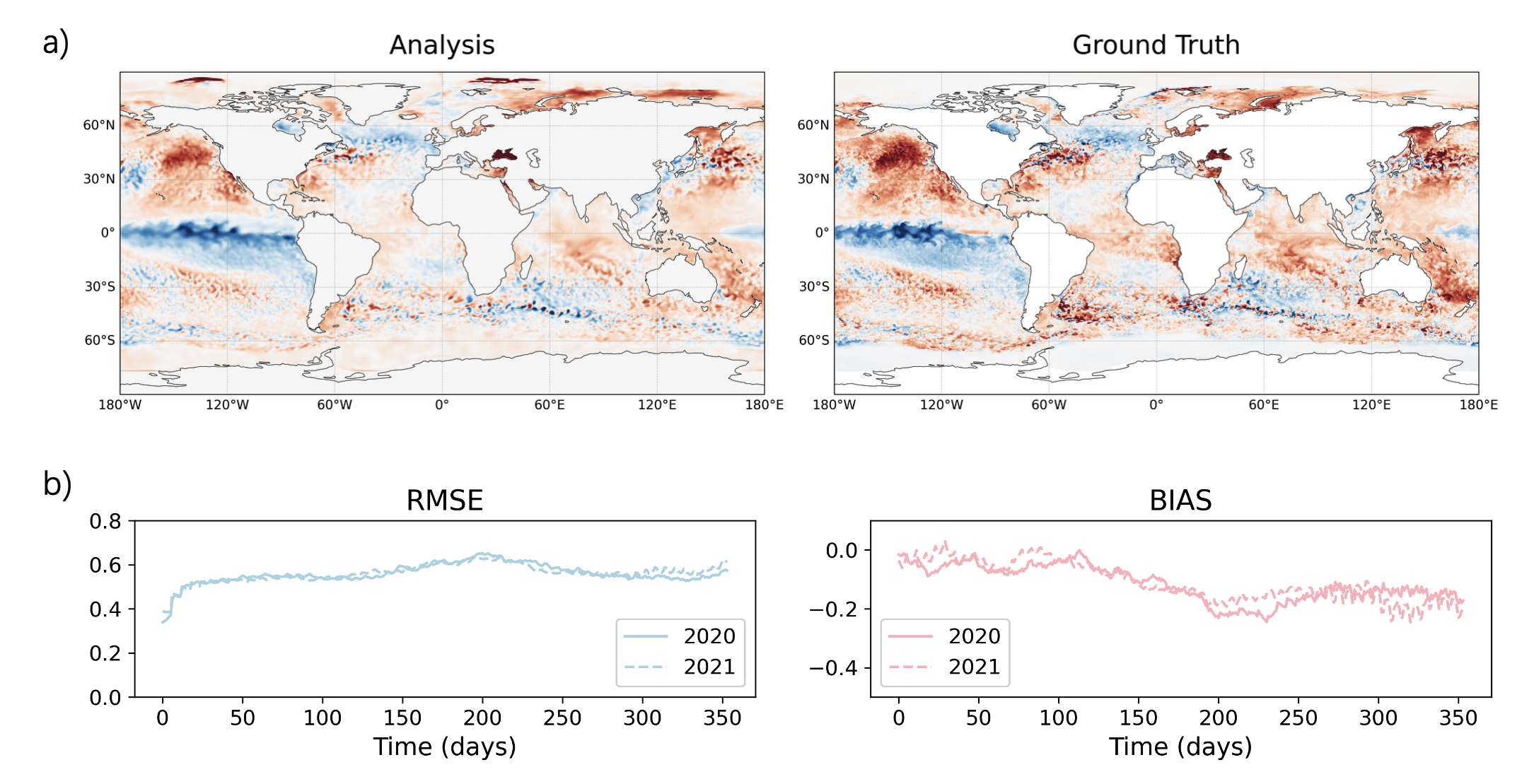}
\caption{End-to-end assimilation results. The panels compare an Ocean-E2E analysis field with the ground truth after 300 assimilation days in 2020 and report the temporal evolution of RMSE and BIAS.}
\Description{Assimilation snapshots and line plots showing that RMSE and BIAS increase initially and then stabilize during the year-long assimilation period.}
\label{fig:assimilation}
\end{figure}

\textbf{Assimilation stability.}
Figure~\ref{fig:assimilation} shows the year-long end-to-end assimilation behavior. After 300 assimilation days, the analysis field preserves major basin-scale warm and cold anomaly patterns, while RMSE and BIAS rise initially and then stabilize during the year-long assimilation period.

\textbf{Additional S2S comparison.}
To be specific, we take the S2S analysis field as the ground truth and use it to initialize Ocean-E2E. The atmospheric boundary condition is acquired through OneForecast~\cite{gao2025oneforecast}. As shown in Table~\ref{tab:forecast_s2s_init}, Ocean-E2E still performs better than S2S.

\begin{table}[h]
\caption{Additional comparison with S2S. The average RMSE for global SSTA is reported. The S2S analysis field is used as the ground truth and initialization for Ocean-E2E; lower is better.}
\label{tab:forecast_s2s_init}
\centering
\small
\setlength{\tabcolsep}{4.5pt}
\renewcommand{\arraystretch}{1.08}
\begin{tabular}{@{}lcccc@{}}
\toprule
\multirow{2}{*}{Model} & \multicolumn{4}{c}{Lead time RMSE} \\
\cmidrule(lr){2-5}
 & 14d & 21d & 28d & 35d \\
\midrule
\multicolumn{5}{c}{\textit{Year 2020}} \\
S2S & 0.5249 & 0.5994 & 0.6618 & 0.7086 \\
\rowcolor{gray!10}
Ocean-E2E & \textbf{0.4321} & \textbf{0.5242} & \textbf{0.5891} & \textbf{0.6304} \\
Gain & 17.8\% & 12.5\% & 11.0\% & 11.0\% \\
\midrule
\multicolumn{5}{c}{\textit{Year 2021}} \\
S2S & 0.5175 & 0.5703 & 0.6121 & 0.6448 \\
\rowcolor{gray!10}
Ocean-E2E & \textbf{0.4350} & \textbf{0.5167} & \textbf{0.5626} & \textbf{0.6022} \\
Gain & 15.9\% & 9.4\% & 8.1\% & 6.6\% \\
\bottomrule
\end{tabular}
\end{table}

\end{document}